\documentclass[%
 aip,
rsi,%
 amsmath,amssymb,
 reprint,%
 floatfix,
]{revtex4-1}

\usepackage{color}
\usepackage{graphicx}% Include figure files
\usepackage{dcolumn}% Align table columns on decimal point
\usepackage{bm}% bold math
\usepackage{multirow}
\newcommand{\fref}[1]{Figure~\ref{#1}}
\newcommand{\tref}[1]{Table~\ref{#1}}
\newcommand{\eref}[1]{Eq.~\eqref{#1}}
\newcommand{\smtd}{{\overline{\mathbf s}}}
\newcommand{\changes}[1]{{\color{black}{#1}}}
\begin{document}

%TODO
\title{Boosting the Conformational Sampling by Combining Replica Exchange with Solute Tempering and Well-Sliced Metadynamics}
%\author{Anji Babu Kapakayala$^{\dagger,\ddagger}$ and Nisanth N. Nair}
\author{Anji Babu Kapakayala}
\affiliation{Department of Chemistry, Indian Institute of Technology Kanpur, Kanpur 208016, India}
\affiliation{School of Pharmacy and Biomedical Sciences, Curtin University, Perth WA 6845, Australia}
\author{Nisanth N. Nair}
\affiliation{Department of Chemistry, Indian Institute of Technology Kanpur, Kanpur 208016, India}
\email{nnair@iitk.ac.in}
%
%
%\affiliation[2]{School of Pharmacy and Biomedical Sciences, Curtin Health Innovation Research Institute and Curtin Institute for Computation, Curtin University, GPO Box U1987, Perth WA 6845, Australia}
%
%\title[Protein folding]{Improving the conformational sampling for the protein folding by combining Replica Exchange with Solute Tempering (REST2) and Well-sliced Metadynamics(WS-MTD) }

%\title[Protein folding]{Exploring free-energy landscape of Trp-Cage folding in explicit solvent by a novel hybrid enhanced sampling approach}

%\begin{document}
\date{\today}
 
%\section{{\bf{Abstract:}}}
\begin{abstract}
Methods that combine collective variable (CV) based enhanced sampling and global tempering approaches are \changes{used in speeding-up} the conformational sampling \changes{and free energy calculation} of large and soft systems with a plethora of energy minima.
%, and to aid in achieving quick convergence in free energy estimates.  
%
In this paper, a new method of this kind is proposed in which the well-sliced metadynamics approach (WSMTD) is united with Replica Exchange with Solute Tempering (REST2) method.
WSMTD employs a divide-and-conquer strategy wherein high-dimensional slices of a free energy surface are independently sampled and combined.
The method enables one to accomplish a controlled exploration of the CV-space with a restraining bias as in umbrella sampling (US), and enhance-sampling of one or more orthogonal CVs using a metadynamics (MTD) like bias.
%
%
%Further, other slow degrees of motion are accelerated by REST2.
\changes{The new hybrid method proposed here enables boosting the sampling of more slow degrees of 
freedom in WSMTD simulations, without the need to specify associated CVs, through a replica exchange scheme within the framework of REST2}.
The high-dimensional slices of the probability distributions of CVs computed from the united WSMTD and REST2 simulations are subsequently combined using the weighted histogram analysis method (WHAM) to obtain the free energy surface.
We show that the new method proposed here is accurate\changes{,} improve\changes{s} the conformational sampling\changes{, and achieves quick} convergence in free energy estimates.
We demonstrate this by computing the conformational free energy landscapes of solvated alanine tripeptide and Trp-cage mini protein in explicit water. 
\end{abstract}
\maketitle

\section{Introduction}
Molecular dynamics (MD) simulations are widely used to study conformational sampling of large biological systems, compute free energetics and identify the mechanism of biochemical processes.\cite{Karplus_McCommon:Nat:2002,Schulten:2009,Dshaw:Science:2010,Shaw:ARB:2012}
%Karlpus_McCammon:Nat:2002,Schulten:Structue:2009,Shaw:ARB:2012} %DOI: https://doi.org/10.1038/nsb0902-646, DOI: 10.1016/j.str.2009.09.001 ,  DOI: 10.1146/annurev-biophys-042910-155245 
%
Modelling large biological systems pose several challenges, primarily due to large number of conformational states and significant energetic and entropic barriers separating different conformational states.\cite{Schulten:2009} %TODO: DOI: 10.1126/science.1187409 
Advanced enhanced sampling techniques are quintessential for achieving proper sampling of conformational states and obtaining reliable free energy estimates.\cite{Chipot:07,Vanden:2009:jcc,Tuckerman:Book,vanGunsteren:JCC:Rev:2010,Ciccotti:12,Giovanni:Entropy:2014,Pratyush:2016,McCammon:2016,Pietrucci:2017,Peters:Book,shalini:review:2019,vashisth:review:2019}
%ADD REFERENCES (order them chronologically: C.AbramsandG.Bussi,Entropy16(1),163–199(2013).
%doi: 10.1080/08927022.2015.1121541
%DOI: https://doi.org/10.1016/j.revip.2017.05.001 [added at 10:30 pm]
%
It is a common practice to use either CV-based or global tempering-based enhanced sampling techniques in combination with MD simulations to improve the sampling.
These approaches are applied to a variety of problems including protein  folding/unfolding, protein-drug binding/unbinding, transport of molecules through membranes, protein aggregation etc.\cite{Adam:AR:2007,Daniel:AR:2011,Wingbermuhle:2020,Acharya:2021}
%TODO1:TODO2}
%DOI: 10.1146/annurev.physchem.58.032806.104614
%DOI: 10.1146/annurev-biophys-042910-155255 

%However conventional MD is suffering from so called time scale issue, where visiting the high barrier crossing event is considered as rare event. To overcome this time scale problem, enhanced sampling techniques are widely used. 

Biased sampling,
generalized-ensemble or global tempering, 
and the combination of \changes{the} both form the three major classes of enhanced sampling methods.
Generalized-ensemble methods achieve a random walk in configurational space by accelerating all the degrees of the freedom either by increasing the temperature of the system or by scaling-down the potential energy. 
Several methods in this category use the replica exchange molecular dynamics (REMD)\cite{remd:1999} algorithm.
\changes{Methods such as} parallel tempering\cite{remd:1999},  replica exchange solute tempering (REST),\cite{rest:2005} and the modified version of REST, called REST2\cite{rest2:2011}, belong to this family. 
Recently modified versions of REST method have been proposed for improving the sampling.\cite{Sugita:JCP:2018,REHT:2021}
Some other distinct \changes{global tempering methods are}
%Stimulated Annealing,\cite{Stimulated_Annealing:1983} Stimulated tempering,\cite{Stimulated_tempering:1992} Wang-Landau,\cite{Wang-Landau:2001} 
accelerated molecular dynamics(aMD),\cite{amd:2004}
chemical flooding\cite{chemical_flooding:1995}, integrated tempering sampling.\cite{ITS:2004}
and accelerated weight histogram method (AWH).\cite{AWH:2014,AWH-1:2021}
%\textcolor{red}{...{\tt TODO: write other methods of this category}}
%

In other family of methods, 
enhanced sampling of certain geometric variables
 of the system, called collective variables (CVs), are carried out by adding a bias potential along CVs or by enhancing the temperature of CVs. 
Umbrella sampling (US),\cite{us:1974,us:1977} metadynamics (MTD),\cite{mtd:2002} adaptive biasing force method\cite{Darve:2001}, logarithmic mean-force dynamics (LogMFD),\cite{LogMFD:2012} driven-Adiabatic Free Energy Dynamics/Temperature Accelerated Molecular Dynamics (d-AFED/TAMD)\cite{tamd:1,tamd:2}, Variational Enhanced Sampling\cite{Variational:1}
%https://journals.aps.org/prl/abstract/10.1103/PhysRevLett.113.090601
and
other variants\cite{wt-mtd:2008,be:mtd:1,pbmtd:2015,shalini:2016} are some examples of biased sampling methods. 
%TODO: add Morishita et al., Phys. Rev. E 2012, 85, 066702
%
%{\tt TODO: add more refs. See Shalini's review and grab references from there}
%
Such methods require {\em a priori} selection of a set of CVs that describes the process of interest. 
The accuracy of the properties computed from the explored conformational ensemble may depend on the quality of the chosen CVs.
Identification of optimal CVs is a challenging task.\cite{Frank_Noe:2017,Henkelman:2017,VPande:2018,Gerhard:2018}
%Ref: https://doi.org/10.1016/j.sbi.2017.02.006
%https://doi.org/10.1021/jacs.7b12191
%DOI 10.1063/1.5007080
%DOI  10.1063/1.5049637 
%
Inclusion of large number of CVs for biased sampling is often required for an efficient exploration of the CV-space and for quick convergence in free energy estimates.
The efficiency of most of the aforementioned
biased-sampling techniques diminishes
with increasing dimensionality of the CV-space.
Sampling of high dimensional free energy landscapes requires advanced techniques such as 
d-AFED/TAMD,\cite{tamd:1,tamd:2}
bias-exchange MTD,\cite{be:mtd:1}
parallel-bias MTD,\cite{pbmtd:2015}
unified-free energy dynamics,\cite{ufed:2012} 
and
temperature accelerated sliced sampling\cite{tass:2017}.

To take the best out of both the generalized ensemble and the CV based biased-sampling methods, hybrid sampling algorithms are proposed.
Replica exchange umbrella sampling (REUS),\cite{reus:2013} 
%Parallel tempering metadynamics (PTMTD),\cite{ptmtd:2006} Bias exchange metadynamics (BE-MTD)\cite{be:mtd:1,be:mtd:2} and 
replica exchange with CV tempering,\cite{rect:2015}
combination of parallel tempering with MTD,\cite{ptmtd:2006,reus-rest}
%combination of REUS with replica exchange with solute tempering,\cite{reus-rest}
replica state exchange MTD,\cite{rse-mtd:2015} and multi-scale sampling using temperature accelerated and REMD\cite{mustar-MD} are examples of such methods.
%
%{\tt \color{red} some more methods: WHAT ARE THESE METHODS?...WRITE LIKE BEFORE}\cite{mustar-MD,reus-rest} 
It has been observed that for modelling protein folding/unfolding and protein-drug binding/unbinding, these methods are advantageous.\cite{ptmtd:2006,rect:2015,reus-rest,reus:2013,Applications:1,Application:2, Application:3,Application:4,Application:REDS:2021}
%\textcolor{red}{TODO}
%
%TODO: this will be added while explaining the REST2 method in the methods section.
%
%REMD is widely used for such study, but this approach hampers with huge computational cost as number of replicas to be simulated increases as $O(N^{\frac{1}{2}})$, where N is the number of degrees of freedom.\cite{fukunishi:2002} To elevate this issue, Berne et. al. introduced replica exchange with solute tempering (REST2)\cite{rest2:2011} which scales the selected Hamiltonian and excludes the solvent interaction during exchange, therefore drastically decreases the replicas needed to simulate. 
%

In this paper, we introduce a new method called globally accelerated sliced sampling (GASS) by integrating the WSMTD and REST2 approach. 
WSMTD\cite{shalini:2016} was introduced by our group for increasing the efficiency 
of MTD method by driving the bias along a specific direction and by controlling the span of the explored CV-space. 
\changes{This method is suited to study systems with broad, flat and unbound free energy landscapes.}
This method has been applied to study various problems\cite{Application:wsmtd:2017,Application:wsmtd:2018_1,Application:2018:2,Application:wsmtd:2020:1,Application:wsmtd:2020:2,Application:wsmtd:2021} and has also been extended to deal with high-dimensional CV-space.\cite{tass:2017}
By combining REST2 with WSMTD, we hope to boost the sampling of hidden transverse coordinates while \changes{exploring} the relevant CV space.
The controlled biased sampling feature of WSMTD could thus be extended to the conformational sampling of large biomolecular systems in solution and is expected to be beneficial for studying problems like protein folding.

\changes{Here, we first} introduce the theory behind the GASS method. 
Subsequently, we present the results of two applications using the GASS method: computation of \changes{conformational} free energy landscapes for (a) alanine tripeptide, and (b) Trp-cage mini protein in water.
%
%In this approach for umbrella window h, REST2 with MTD on selected CVs will be employed. The unbiased probabilities will be combined using Weighted histogram analysis method (WHAM)\cite{wham:1,wham:2} after re-weighting MDT bias. Consequently free energy will be reconstructed along selected CVs. We refer to the introduced approach as {\bf{Globally Accelerated Sliced Sampling (GASS)}}. The method is first tested on alanine tri peptide in explicit water and then applied to folding of Trp-cage mini protein in explicit water. The performance of standard REST2 and GASS are carefully compared. We show that with GASS it is possible to obtain the better conformational sampling and faster convergence in free energy estimates. 
%
\section{Methods}
%\textbf{Well Sliced Metadynamics (WS-MTD)}\\
%First, we summarize the well sliced metadynamics (WS-MTD) methodology according to the original work.\cite{shalini:2016} 
% Therefore this method takes an advantage over conventional methods to study the systems with broad, flat and unbound free energy landscapes.

\changes{WSMTD is a CV based enhance sampling method
which can help to achieve a controlled exploration of the CV space by combining US and MTD.
While US is ideal for a controlled or directional sampling along one CV, MTD is advantageous in sampling orthogonal CVs in a self guided manner. 
In WSMTD, our interest is in computing an $n$-dimensional free energy surface $F(\mathbf s$) as a function of CVs, $\mathbf s(\mathbf R)\equiv \{s_1,\cdots,s_n\}$, where $\mathbf R$ is the set of atomic coordinates.
%
%The target is obtain a controlled exploration along the CV $s_1$ is anticipated. 
%
}
WSMTD\cite{shalini:2016} uses the Lagrangian, 
%WS-MTD samples the CV space $\bf{s}=(\bf{s}_1,\bf{s}_2)$ by carrying the $M$ independent MTD simulations with $M$ umbrella potentials along the coordinate $s_1$ using the following biased Hamiltonian form.
%\begin{widetext}
\begin{eqnarray}
      \mathcal L_h^{\rm ws}(\mathbf R ,\dot {\mathbf R}) 
   = \mathcal L^{0}(\mathbf R ,\dot {\mathbf R}) - 
      W^{\rm b}_{h}(s_1) 
      - V^{\rm b}(\smtd,t), 
%      \enspace \enspace h=1, ..., M \enspace  
      \label{e:wsmtd:Lagrangian}
\end{eqnarray}
%\end{widetext}
$h=1, ..., M $,
where, $\mathcal L^0$ is the unbiased Lagrangian, \changes{$\dot{\mathbf R}$
is the set of velocities, and $\smtd \equiv \left \{ s_2,\cdots,s_n  \right \}$.}

\changes{In} WSMTD\changes{, we apply a} restraining bias
\begin{eqnarray}
    W^{\rm b}_{h}(s_1) = \frac{1}{2}k_{h}(s_1 - z_h)^2
    \label{e:us:bias}
\end{eqnarray}
placed at $s_1 = z_h$\changes{. T}he parameter $k$ determines the curvature of the biasing potential at $h^{\rm th}$ window.
\changes{Here, $M$ umbrella windows placed from $z_1$ to $z_M$ along  $s_1$ define the span of sampling of that CV.}
This helps in achieving a controlled sampling of the coordinate $s_1$, similar to the conventional US technique.
\changes{In WSMTD, we also apply} a well-tempered MTD bias potential
%$V^{\rm b}(\smtd,t)$, as
\begin{equation}
V^{\rm b}( \smtd,t)= \sum_{\tau < t}w_\tau \exp\left[- \frac{ \left ( \smtd - \smtd(\tau) \right )^2}{2(\delta s)^2}  \right] \enspace ,
\end{equation}
in order to sample the orthogonal coordinates $\smtd$.
Here, $\delta s$ is the width parameter defining the Gaussian potential and  
%
%In Well Tempered MTD(WT-MTD)\cite{wt-mtd:2008} the height of the Gaussian posited is scaled as 
%
\begin{equation}
    w_\tau = w_0 \exp \left[- \frac{V^{\rm b}(\smtd,t)}{k_{\rm B} \Delta T}  \right] \enspace , 
\end{equation}
\changes{Where} $w_0$ is the Gaussian height parameter, $k_{\rm B}$ is the Boltzmann constant, and $\Delta T$ is a tempering parameter.

The free energy surface, $F(\mathbf s)$, is reconstructed from the probability distribution, $P(\mathbf s)$, of the CVs as,
\begin{equation}
    F(\mathbf s)=-\frac{1}{\beta}\ln P(\mathbf s)
\end{equation}
where $\beta = (k_{\rm B} T)^{-1}$. \changes{In order to obtain $P(\mathbf s)$, a} time independent probability distribution, $P_h^{\rm u}(\mathbf s)$, is obtained by reweighting the MTD time-dependent bias potential as,
\begin{widetext}
\begin{eqnarray}
    P^{\rm u}_{h}(\mathbf s^\prime)=
    \frac{\int_{t_{\rm min}}^{t_{\rm max}}d\tau \, \exp \left[ 
    \beta \{V^{\rm b}(\smtd,\tau)-c(\tau)\} \right]  
    \prod_{\alpha=1}^n \delta(s_\alpha(\tau)-s_\alpha^\prime) 
    }
    { \int_{t_{\rm min}}^{t_{\rm max}}d\tau \, \exp \left[ 
    \beta \{V^{\rm b}(\smtd,\tau)-c(\tau)\} \right]  
    }  \enspace  , %\enspace \changes{h= 1,\cdots,M}
    \label{e:prob:unbias:mtd}
\end{eqnarray}
\end{widetext}
\changes{for each umbrella window $h$}.
\changes{In the above,} $c(t)$ is evaluated as\cite{Tiwary-Parinello:2015}
\begin{equation}
    c(t)=\frac{1}{\beta}\ln \left [ \frac{\int d\smtd \exp[-\beta \gamma V^{\rm b}(\smtd,t)] }{\int d \smtd \exp[-\beta (\gamma-1)  V^{\rm b}(\smtd,t)]}\right ]
\end{equation}
and $\gamma = {(T+\Delta T)}/{\Delta T}$ .
%
%{\tt \color{red} USE THE C(T) EQUATION IN SHALINI'S JCP PAPER....} 
%\begin{equation}
   %\color{red} c(t)=\frac{1}{\beta}\ln \left \{ \frac{\int %d\mathbf s \,  \exp \left [-\beta F(\mathbf s) \right %]}{\int d \mathbf s \exp \left [-\beta F(\mathbf %s)+V^{\rm b}({ { \mathbf s}},t) \right ]} \right \}   %\enspace . 
%\end{equation}
%NN: I am removing this, as we are substituing this directly in c(t) equation:
%
%However $F(s)$ in Eq. (6) is the time-independent free energy, which can be computed using the Tiwary-Parrinello time-independent free energy estimator\cite{Tiwary-Parinello:2015} as
%\begin{equation}
%    F(s)=-\alpha V^b(s,t)+ \frac{1}{\beta}\ln \int ds \exp[ \alpha \beta V^b(s,t)]
%\end{equation}
%In the above,  the distribution $P^{\rm u}_h(\mathbf s)$ is not reweighted for the restraining bias.
%
Now we employ the weighted histogram analysis method (WHAM)\cite{wham:1,wham:2} to combine the distributions $P^{\rm u}_h(\mathbf s)$, $h=1,\cdots,M$ and reweight the restraining bias by self-consistent calculations using, 
%
%{\tt \color{red} TODO: PLEASE CHECK THE FOLLOWING EQUATIONS..I THINK THERE IS SOME MISTAKE}
%
% umbrella potential and combine this with WHAM such that slices of probability densities $P^{u}(s_1,s_2)$, $h=1, ..., M$ can be joined to obtain the unbiased distribution $P(s_1,s_2)$ by self-consistently solving WHAM equations\cite{wham:1,wham:2}
\begin{equation}
P(\mathbf s) = \frac{\sum^{M}_{h=1}n_h P^{\rm u}_h(\mathbf s)}{\sum_{h=1}^{M}n_h \exp [\beta f_h] \exp[-\beta W_h^{\rm b}(s_1) ]} 
\end{equation}
and
\begin{equation}
    \exp [-\beta f_h]=\int d \mathbf s \,  \exp \left [-\beta W_h^{\rm b}(s_1)] \, P(\mathbf s) \right ] \enspace .
\end{equation}
\changes{In the above,} $n_h$ is the number of configurations sampled in the $h^{\rm th}$ \changes{umbrella} window. 
%Finally, the free energy surface $F(s_1,s_2)$ is constructed using
%\begin{equation}
%    F(s_1,s_2)=-\frac{1}{\beta}\ln P(s_1,s_2)
%\end{equation}
%
%\newpage
%\noindent\textbf{Replica Exchange Solute Scaling (REST2)}\\

In the REST2 method,\cite{rest:2005,rest2:2011} the potential energies of a selected set of atoms in each replica are scaled-down by some parameter $\lambda$ to enhance their sampling, and conformations sampled in these scaled-replica are exchanged with the non-scaled replica to improve the conformational sampling of the latter.
%
%
%This would increase the effective temperature of the system involving these atomswhich enhances the sampling by increasing the effective temperature of the system.
%
The atoms which are part of the scaled-potential terms (in a molecular mechanics force-field) are referred to be in ``hot'' region and the rest \changes{are said to be} in ``cold'' region.
%, where the scaled part becomes the hot region and the unscaled part becomes the cold region. 
%In classical simulations the potential energy of the system is constructed from defined force field parameters. 
Pair-wise potential form of the molecular mechanics force-field
makes it easier to selectively scale potential contributions for a set of atoms and 
differently scale torsional, electrostatics, and Lennard-Jones terms.
%
%Therefore, the force field parameters that contribute to the potential energy barriers of the hot region (i.e. dihedral angles and electrostatics and Lennard-Jones interactions) are scaled by some factor $\lambda$. 
Typically, for a solvated protein system, the potential energy of a replica $m$ is computed from the modified 
contributions of the protein-protein (pp), protein-water (pw) and water-water (ww) interactions, as 
\begin{equation*}
    U_m^{\rm REST2}({\bf{R}})= \lambda_m \, U_{\rm pp}({\bf{R}})+\sqrt{\lambda_m} \, U_{\rm pw}({\bf{R}})+U_{\rm ww}({\bf{R}}), 
    %\enspace \enspace 
\end{equation*}
$ m=0,\cdots,N_{\rm r}-1$,
where 
$\lambda_m={\beta_m}/{\beta_0}$, %$\frac{\beta_m}{\beta_0}=\lambda$, Equation (11) can be written as
%\begin{equation}
%    E_m^{REST2}({\bf{R}})= E_{pp}({\bf{R}})+\sqrt \lambda E_{pw}({\bf{R}})+E_{ww}({\bf{R}})
%\end{equation}
$N_{\rm r}$ is the number of replicas, and $\beta_m > \beta_0$.
 The $\beta_m$ parameter is \changes{set} different for different replicas.
For $m=0$, i.e., for the unscaled replica, we take $\beta_0 = (k_{\rm B} T_0)^{-1}$, where $T_0$ is the
physical temperature of the system.
%
%In a REST2 simulation, several replicas are run simultaneously and independently, with the modified potentials corresponding to the chosen $\lambda$ values. At certain time intervals, %
Exchange between adjacent replicas is attempted by swapping their atomic coordinates based on the Metropolis exchange criterion     
\begin{eqnarray} p(i\rightarrow j)=\min(1,e^{-\Delta_{i,j}}) \enspace, \label{e:detailed:balance}
\end{eqnarray}
\changes{with}
\begin{widetext}
\begin{equation}
\Delta_{i,j}=(\beta_i - \beta_j)\left [ (U_{\rm pp}({{\mathbf R}_j})- U_{\rm pp}({{\mathbf R}_i})) + \frac{\sqrt{\beta_0}}{\sqrt{\beta_i}+\sqrt{\beta_j}}(U_{\rm pw}({\mathbf{R}_j})- U_{\rm pw}({\mathbf{R}_i}))      \right ] 
\enspace .
\label{e:rest2:delta}
\end{equation}
\end{widetext}
%{\tt \color{red} I THINK THE ABOVE EQUATION HAS SOME ERROR. SEE THE SQUARE ROOT OVER BETA I IN THE DENOMINATOR. In the above, is it -$\Delta$ or $+\Delta$?}
%
%The solvent-solvent interaction terms are cancelled out in the expression of exchange probability. 
The absence of solvent-solvent interaction in the above expression boosts the acceptance  between neighbour replicas \changes{compared to conventional parallel tempering simulations}.
%compared to a typical replica-exchange method.
%\textcolor{red}{Add REST2 picture: Draw good one}
%\newpage
%\\
%\\
%\noindent\textbf{Globally Accelerated Sliced Sampling (GASS)}\\

%
\changes{WSMTD is usually used for cases where  the number
of CVs is small.
To improve the efficiency of WSMTD for large number of CVs, TASS method was proposed.\cite{tass:2017}
Like WSMTD, TASS is also a CV-based enhanced sampling method. 
However, it is nontrivial to find suitable CVs for enhancing the slow global motions in large soft matter systems like solvated proteins.\cite{Bolhuis:BPJ:10}
As discussed earlier, global tempering methods like REST2 can achieve this by potential energy scaling in replica exchange.
Thus to allow enhanced sampling of global motions of large molecular systems in WSMTD, we introduce the GASS approach by combining WSMTD and REST2.
}

\changes{Like in WSMTD, our aim in GASS simulations is to compute the free energy surface $F(s_1,\cdots,s_n)$ where $n$ is typically not more than three. 
We 
prefer to achieve a controlled sampling along
$s_1$ together with a self-guided biased sampling along
other $n-1$ CVs.
For a quick and accurate estimation of $F(\mathbf s)$, we also intent to boost the enhance sampling of slow
global conformational changes without defining
 additional CVs.}
In GASS approach, we use the WSMTD Lagrangian (\eref{e:wsmtd:Lagrangian}) for the windows $h=1,\cdots,M$.
For every umbrella $h$, we perform REST2 simulation considering $N_{\rm r}$ replicas, and construct the  probability 
density $P^{\rm u}_h(\mathbf s)$ from the REST2 trajectory for the $m=0$ replica using \eref{e:prob:unbias:mtd}.
%at effective temperature $\beta_0$.
%
%sample orthogonal coordinates more efficiently and reconstruct the free energy surface.
%\\{\textbf{Derivation of GASS:}}\\

In order to accommodate the bias-potentials acting on the replica for a given window $h$, 
a modified equation for computing exchange probability is used for GASS.
To satisfy the condition of detailed balance  we use
\begin{equation}
    \Delta_{i,j}= \Delta^{(1)}_{i,j} + \Delta^{(2)}_{i,j}
\end{equation}
in \eref{e:detailed:balance}, where $\Delta^{(1)}_{i,j}$ is given by \eref{e:rest2:delta} and
%with
%\begin{eqnarray}
%\Delta^{(1)}_{i,j} = \beta_0 \left\{ \left [U_{i}(\mathbf R_{j})-U_{i}(\mathbf R_{i})]-[U_{j}(\mathbf R_{j})-\mathbf U_{j}(\mathbf R_{i}) \right ] \right\} \enspace  , \mathrm{and}
%\end{eqnarray}
\begin{eqnarray*}
\Delta^{(2)}_{i,j} = \beta_0 \left\{ \left [V^{\rm b}_{i}(\smtd_{j}, t) - V_{i}^{\rm b}(\smtd_{i},t)]-[V_{j}^{\rm b}(\smtd_{j},t) - V_{j}^{\rm b}(\smtd_{i},t)  \right ] \right\} \enspace .
\end{eqnarray*}
 Here $i$ and $j$
are the replica indices; $\smtd_i$ specifies the set of CVs  $(s_2,\cdots,s_n)$ corresponding to the coordinates $\mathbf R_i$ of the replica $i$.
%
%We consider  $j=i+1$, for $i=0,\cdots,N_{\rm r}-2$, and $\beta \equiv \beta_0$.
%any window $h$ and time $t$.
%
%
%For a an umbrella window $h$,
% \begin{equation}
%     H_h({\bf{R,P,s,\Dot{s}}})=H_h^{0}({\bf{R,P}})+W_h^{b}(s_1)+V_h^{b}(s_2,t%)
% \end{equation}
%The potential energy of the Replica $k$ at $\lambda_k$ can be written as ,
%\begin{equation}
%     U_{k}({\bf{R,s}})=U_{k}^{0}({\bf{R}})+W_k^{b}(s_1)+V_k^{b}(s_2,t)
% \end{equation}
% The potential energy of the Replica $k+1$ at $\lambda_{k+1}$ can be written as ,
%\begin{equation}
%     U_{k+1}({\bf{R,s}})=U_{k+1}^{0}({\bf{R}})+W_{k+1}^{b}(s_1)+V_{k+1}^{b}(s_2,t)
% \end{equation}
%The exchange probability as per metropolis criteria is,
%\begin{equation}
%    w=e^{-\Delta_{k,k+1}}
%\end{equation}
%Where,
%\begin{equation}
 %   \Delta_{k,k+1}=\beta \left\{[U_k(R^{k+1},s^{k+1})-U_k(R^{k},s^{k})]-[U_{k+1}(R^{k+1},s^{k+1})-U_{k+1}(R^{k},s^{k})]          \right\}
%\end{equation}
%Substitute Eq. (15) into Eq. (17) then split for $\Delta$ ,
%
%$$\Delta = \Delta_1+\Delta_2+\Delta_3$$
%
%Then we can write as,
%\begin{equation}
%    \Delta_1=\beta \left\{[U_{0,k}(R^{k+1})-U_{0,k}(R^{k})]-[U_{0,k+1}(R^{k+1})-U_{0,k+1}(R^{k})]          \right\}
%\end{equation}
%
%\begin{equation}
%    \Delta_2=\beta \left\{[V_{k}(s^{k+1},t)-V_{k}(s^{k},t)]-[V_{k+1}(s^{k+1},t)-V_{k+1}(s^{k},t)]          \right\}
%\end{equation}
%
%\begin{equation}
%    \Delta_3=\beta \left\{[W_{k}(s^{k+1})-W_{k}(s^{k})]-[W_{k+1}(s^{k+1})-W_{k+1}(s^{k})]          \right\}
%    \end{equation}
%
Since all the replicas for a given window $h$ have the same umbrella bias (\eref{e:us:bias}),
it is not contributing to the calculation of $\Delta_{i,j}$.
\changes{See Supporting Information for a derivation of the above expression for exchange probability. 
}
The GASS method has been implemented using the GROMACS/PLUMED interface.\cite{GROMACS,plumed2:2014} %TODO
%ve
All simulations in this paper were performed using  GROMACS-2018.6\cite{GROMACS} patched with PLUMED-2.2.6\cite{plumed,plumed2:2014} and HREX.\cite{Bussi_rest2:2014} %and other modifications required for the GASS method.

\begin{figure*}[t]
\centering
    \includegraphics[width=0.5\textwidth]{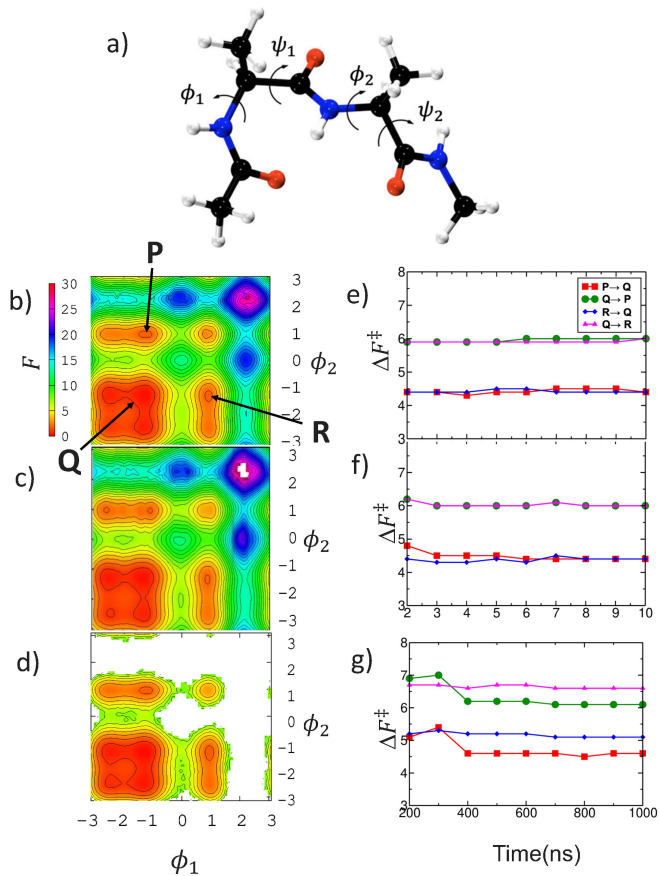}
    \caption{(a) Ball and stick representation of alanine tripeptide (Color codes: H - white, C - black, O - red, and N - blue). 
     Free energy surface  $F(\phi_1, \phi_2)$ 
     computed using 
     GASS (b), WSMTD (c), and REST2 (d) simulations of alanine tripeptide in water are given. 
    Convergence of free energy barriers for $\mathbf P \rightarrow \mathbf Q$, $\mathbf Q \rightarrow \mathbf P$, $\mathbf R \rightarrow \mathbf Q$, and $\mathbf Q \rightarrow \mathbf R$ as a function of simulation time in 
    GASS (e), WSMTD (f), and REST2 (g) simulations are shown.
    Here the angles are in radians and free energies are in kcal~mol$^{-1}$.
    }
    \label{fig:fes_alatripep}
\end{figure*}

\section{Results and Discussion}

The accuracy and efficiency of the GASS method was first investigated for 
the exploration of the  conformational free energy landscape of alanine tripeptide in explicit water.
The performance of the method is compared with REST2 and WSMTD.
Thereafter, it was tested for computing the free energy landscape of unfolding of Trp-cage protein in water.
\subsection{Alanine Tripeptide in Water}

\begin{table*}[htbp]
\centering
\begin{tabular}{||c| c | c | c | c | c | c | c ||} 
 \hline
 Method & Simulation Time (ns) & \multicolumn{4}{|c|}{$\Delta F^\ddagger$} & \multicolumn{2}{|c||}{$\Delta F$} \\ \cline{3-8}
        &                      &  $\mathbf P \rightarrow \mathbf Q$    & $\mathbf Q \rightarrow \mathbf P$ & $\mathbf R \rightarrow \mathbf Q$ & $\mathbf Q \rightarrow \mathbf R$ & $\mathbf P - \mathbf Q$ & $\mathbf R - \mathbf Q$
     \\  
 \hline\hline
 GASS & 10 & 4.4 & 6.0 & 4.4 & 6.0 &   1.6 &   1.6  \\ 
% TASS & 10 & 5.3 & 7.2 & 4.8 & 6.5\\
 WSMTD & 10 & 4.4& 6.0 & 4.4 & 6.0 &   1.6  & 1.6  \\
 REST2 & 1000 & 4.6 & 6.1 & 5.1& 6.6 & 1.5  & 1.5 \\[1ex]
 \hline
\end{tabular}
\caption{Free energy barriers ($\Delta F^\ddagger$)  and free energy differences ($\Delta F$), in kcal~mol$^{-1}$, computed from the free energy landscape $F(\phi_1,\phi_2)$ of alanine tripeptide in water using various methods.}
\label{table:tab_alatripep}
\end{table*}

Alanine tripeptide in water was modeled using {\tt AMBER14SB}\cite{ff14SB} and TIP3P\cite{TIP3P} force-fields.  
We took the amino acid sequence ACE-ALA-ALA-NME where the terminal residues ACE and NME were acetyl and N-methyl amide, respectively, and the total number of water molecules in the system is 785. 
All the bonds in the system were constrained and the equations of motion were integrated using a time step of 1~fs. 
Long range electrostatics was treated using the particle-mesh Ewald technique, as available in GROMACS.\cite{PME} 
Temperature of the system was controlled using the stochastic velocity rescaling thermostat.\cite{v-scaling:thermostat} 
Density was converged in 100~ps of $NPT$ simulation and the 
equilibrated cell volume \changes{achieved was} $34 \times 36 \times 30 $~{\AA}$^{3}$.
GASS simulations were then carried out in the $NVT$ ensemble using the equilibrated density.
In REST2, all the peptide atoms were taken in the hot region and 5 replicas were considered with the values of $\lambda_m$ ranging from 1.0 to 0.3 following a geometric distribution. 
Exchanges between the neighboring replicas were attempted every 1000 MD steps.
%, which provided the 50\% of exchange probability.
 The two Ramachandran angles $\phi_1$ and $\phi_2$ were chosen as CVs for biasing; see \fref{fig:fes_alatripep}a. 
Umbrella bias was applied along $\phi_1$ from $-\pi$ to $+\pi$ at an interval of 0.2 radians with $\kappa_h=1.2 \times 10^{2}$ kcal~mol$^{-1}$ rad$^{-2}$ and \changes{a well-tempered MTD} bias was applied along $\phi_2$. 
The MTD bias was updated every 500~fs, and the bias parameters $w_0=0.6$~kcal~mol$^{-1}$,  $\delta s=0.05$~radians, and $\Delta T = 900~$K were taken. 
%
%GASS simulations were performed independently for each umbrella window. 
\changes{It is noted that the choice of types of biases} 
%in passing 
%that the choice of the type of bias 
applied on \changes{$\phi_1$ and $\phi_2$} 
%CVs 
\changes{were arbitrary}. 

%
%and the final results should not affect by this selection.

%
%A total of 33 umbrella windows were considered along $\phi_1$ from $-\pi$ to $\pi$, with an interval of 0.2 radians. 
%
For demonstrating the performance of GASS, we carried out independent WSMTD, and REST2 simulations with identical setups. %
%WS-MTD and TASS were run for 10~ns per window, whereas REST2 simulation was performed for 1000~ns. 
%
%Technical detail for these calculations were identical to that of GASS.
%

   %Here the symbols $\blacksquare, \bullet,
    %\blacklozenge,$ and $\blacktriangle$ are used for free energy barriers for $P \rightarrow Q, Q \rightarrow P, R \rightarrow Q$, and $ Q \rightarrow R,$ respectively.
%    {\tt \color{red} TODO: make changes as suggested in SKYPE message box}}
 
 Figure ~\ref{fig:fes_alatripep} shows the free energy surface $F(\phi_1,\phi_2)$ computed using GASS, WSMTD, and REST2 and the convergence of free energy barriers for these  surfaces.
%
%
%{\textbf{Discussion:}}\\
%After the careful analysis of GASS trajectories , we measured the convergence of free energy barriers by comparing the free energy surfaces obtained for various simulation lengths.
%
The free energy barriers $\mathbf Q \rightarrow \mathbf P$ and $\mathbf Q \rightarrow \mathbf R$ 
are expected to be nearly the same due to the symmetry, and the same holds true for the corresponding reverse barriers.
From \fref{fig:fes_alatripep}e, it is clear that very accurate predictions of free energy barriers are possible even after 2~ns of GASS simulation.
As anticipated, the $\mathbf Q \rightarrow \mathbf P$ and  $\mathbf Q \rightarrow \mathbf R$ barriers are found to be equal, and the same is true for $\mathbf P \rightarrow \mathbf Q$ and $\mathbf R \rightarrow \mathbf Q$ barriers.
A smooth free energy surface was obtained after 10~ns of GASS simulation, and the system  swept through the entire two-dimensional landscape.

%We have also performed WSMTD (without coupling to REST2) and the free energy surface $F(\phi_1,\phi_2)$ was also computed.
%from these simulations to compare the performance of GASS with WSMTD.
%We also performed independent WSMTD and computed the free energy surface.
The free energy barriers computed from \changes{independent} WSMTD are agreeing very well with the GASS results.
Interestingly, the convergence of free energy barriers \changes{in their runs} was as quick as that \changes{using} GASS.
However, the higher energy region\changes{s near } $\phi_1=2.4$, $\phi_2=2.4$ (radians) \changes{were} not well explored in WSMTD (\fref{fig:fes_alatripep}c).
%, while it was better with GASS.
%
The \changes{results of independent } REST2 simulations  (\fref{fig:fes_alatripep}d,g) show that conformations far from the free energy minima are not well explored even after \changes{1~$\mu$s} and the estimates of the free energy barriers \changes{($\Delta F^\ddagger$)} are not well converged.
On the other hand, free energy differences \changes{($\Delta F$)} computed using GASS, WSMTD, and REST2 are 
agreeing very well; see \tref{table:tab_alatripep}.

These results give us the confidence that GASS method is able to \changes{provide accurate free energy estimates and could} achieve quick conformational sampling.
The method is evidently performing better than both WSMTD and REST2 methods.

\subsection{Conformational Landscape of Trp-Cage in Water}

\begin{figure*}[htbp]
\centering
    \includegraphics[width=0.5\textwidth]{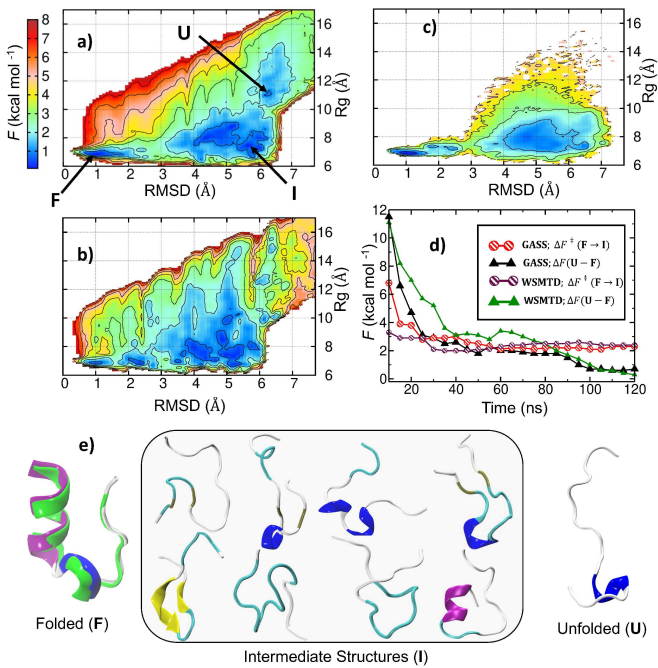}
    \caption{ Conformational free energy landscape of Trp-cage protein computed along the RMSD and Rg CVs using (a) GASS (b) WSMTD (c) REST2. (d) Convergence of free energy difference
    ($\Delta F = F ( {\mathbf{U}} - {\mathbf{F}})$ )
    between the unfolded ({\bf U}) and the folded state ({\bf F}) and convergence of free energy barrier for going from {\bf F} to the intermediate {\bf I}  ($\Delta F^\ddagger ({\mathbf{F}\rightarrow\mathbf{I}})$) are plotted as function of simulation time. 
    %(c) Free energy surface obtained from REST2  simulation after 3~$\mu$s. 
    (e)  Representative conformations of {\bf F}, {\bf I} and {\bf U} states are shown. The NMR structure (green) is overlapped with {\bf F} for a comparison.  Number of other conformations were also observed in {\bf I} and {\bf U} states; See SI Section 4.}
    %{\tt TODO: CHANGE THE FIGURES WITH THE ONES YOU USED FOR YOUR PRESENTATION. CHANGE THE FIGURES AS PER THE MAIN TEXT...SEE BELOW.}}
    \label{fig:fes_trpcage}
\end{figure*}

As next, we investigated the unfolding/folding landscape of Trp-cage in explicit water using the GASS method.
This is an ideal problem to study using GASS as a controlled sampling along a ``folding/unfolding coordinate'' using the restraining bias could drive the conformations from folded to unfolded states, or vice-versa.
At the same time, accelerated sampling of several orthogonal \changes{coordinates} are essential to boost the exploration of the conformational states \changes{for such systems} which can be achieved by \changes{using} MTD bias and REST2.

Trp-\changes{c}age is a 20 amino acid mini-protein (NLYIQ WLKDG GPSSG RPPPS)  designed by Neidigh et al.\cite{trp:2002:original} from 39 amino acid extendin-4 peptide.
It is a fast-folding protein and is \changes{considered as} an ideal model system for testing \changes{computational} methods developed for protein-folding problems. 
It contains a short $\alpha$-helix (residues 2-9), a $3_{10}$-helix (residues 11-14), and a C-terminal polyproline-II helix.
%{\tt \color{red} TODO: I DONT UNDERSTAND THIS? ``to pack against the central tryptophan (Trp-6).''}
%
%A number of computational studies have reported the free energy landscape of folding/unfolding of this protein.\cite{trp:2002,trp:2003,be:mtd:1,be:mtd:2,trp:2011,trp:2012,trp:2013,trp:2014,trp:2015,trp:2015:amd,trp:2017,trp:2018,REHT:2021}
%\cite{TODO:cite-all-important-works}

%
The initial Trp-cage structure for our simulations was built based on the folded NMR structure PDB ID 1L2Y.\cite{trp:2002:original} 
The protein was first solvated in a periodic $80 \times 80 \times 80$~\AA$^3$
TIP3P water box \changes{with} 1~g~cm$^{-3}$ density. 
\changes{We used} AMBER ff14SB\cite{ff14SB} force-field \changes{for the protein.} \changes{L}ong-range electrostatic interactions were evaluated using the Particle Mesh Ewald method.\cite{PME}
All the bonds in the system were constrained using the LINCS algorithm.\cite{LINCS}
A time step of 2~fs was used to integrate the equation of motion. 
Temperature of the system was controlled by stochastic velocity rescaling thermostat.\cite{v-scaling:thermostat}  
A 1~ns long $NPT$ run was initially carried out using Parrinello-Rahman barostat\cite{Parrinello:Rehman} to obtain converged density and the equilibrated cell volume \changes{was found to be} $83 \times 83 \times 83$~{\AA}$^3$.
Using the equilibrated \changes{system}, we performed  10~ns equilibration in $NVT$ ensemble to generate initial structure\changes{s} for the GASS simulation.
%{\tt \color{red} CHECK THIS}
%{\tt \color{red} TODO: WRITE ABOUT INITIAL NPT, CONVERGED CELL PARAMETERS ETC.}
%
%Calculations were performed using GROMACS-2018.6\cite{GROMACS} patched with PLUMED-2.2.6\cite{plumed,plumed2:2014}.
%and integrated with the modifications required for enabling the GASS method. {\tt TODO: i think we have already said this in Theory section - need to repeat! CHECK..AND DELETE}
 %
 
 All the protein atoms were chosen in the hot region for the REST2 \changes{calculations.} 
 We chose 20 replicas with $\lambda_m$ values ranging from 1.0-0.3 to obtain a good exchange probability. 
 The exchange of coordinates was attempted \changes{after} every 1000~MD steps.

We chose to apply restraining bias along the backbone root mean square deviation (RMSD) CV  in order to drive the unfolding of the protein from the folded starting structure; 
See SI~Section~1.1 for the definition of the coordinate.
As the reference structure for computing RMSD, we took the backbone structure of all the heavy atoms \changes{for the residues 1 to 15} in the PDB 1L2Y.
This CV will serve as the ``unfolding coordinate'' which will guide the system from folded state to unfolded state in a controlled manner.
A total of 45 umbrella windows were placed every {0.20}~{\AA}, starting from 0.20~{\AA}.
MTD bias was applied along the radius of gyration (Rg) in all the windows; See SI~Section~1.2 for the definition of Rg.
This bias potential could enhance the conformational sampling by triggering the breakage of $\alpha$-helices, thereby promoting the system to visit unfolded states.
Since GASS approach enables us to have different transverse coordinates for different windows, we 
 chose the $\alpha$-helicity (SI~Section~1.3) CV together with Rg for applying MTD bias for the windows placed between RMSD values 6.0~{\AA} and 8.0~{\AA}.
%
% was essential to obtain the ensemble of unfolded conformations.
%
For the MTD bias, we choose the Gaussian parameters $w_0=0.6$~kcal~mol$^{-1}$, and $\delta s = 0.02$.  
We took $\Delta T= 2700$~K and $5700$~K when we used one-dimensional and two-dimensional MTD biases, respectively.
We scaled-up the width of the Gaussian by a factor 10 in the direction of the $\alpha-$helicity coordinate because of the
large amplitude fluctuation of this CV compared to Rg.
The starting umbrella window at RMSD=$0.20$~{\AA}  pertained to a folded structure, resembl\changes{ing} well with the NMR structure (PDB~ID: 1L2Y). 
%
%We used the equilibrated structure of this window as the initial structure for the neighboring window.
%
Starting structure of all the other windows with higher value of RMSD was taken as the equilibrated structure of the \changes{preceding} window.  

%
%Whereas from umbrella \angstrom{6.00} to \angstrom{8.00} ,2D METAD bias with bias-factor 20.0 was applied along Rg and Alpha-helicity CVs with Gaussian widths 0.02, 0.2 respectively. 2 $\mu s$ REST2 simulation was performed for the comparison.

%\\
%\begin{figure}[htbp]
%\centering
%    \includegraphics[width=0.8\textw%    \caption{Free-energy difference of folded and unfolded states of TRP-Cage protein plotted as a function of simulation time.}
%    \label{fig:convergence_trpcage}
%\end{figure}
%{\textbf{Discussion:}}\\

%Trp-cage is a 20-residue miniprotein, which is believed to be the fastest folder known so far. 
%
%In this study, the folding free energy landscape of Trp-cage has been explored in explicit solvent by using an amber ff14SB force field with periodic boundary condition.
%
%
%Several constructs of increasing stability were made by gradually introducing stabilizing features such as helical N-capping residues and a solvent-exposed salt-bridge. 
%
%

%To compare standard REST2 with GASS in a fair way, we use exactly the same parameters (number of replicas, exchange frequencies etc) in both cases.
%{\textcolor{blue}{have to explain about trp cage protein:}}
%{\textcolor{red}{Many previous studies have been reported on Trp-Cage system.\cite{trp:2002,trp:2003,be:mtd:1,be:mtd:2,trp:2011,trp:2012,trp:2013,trp:2014,trp:2015,trp:2015:amd,trp:2018}
%}}
%
%
We reconstructed the free energy surface in the RMSD-Rg space after 120~ns of the simualation (\fref{fig:fes_trpcage}a). 
Free energy surfaces constructed after 50~ns, 75~ns, 100~ns of GASS simulation are presented in SI~Section~2.
%TODO change 2b to 2c in the figure.
%Move figure 3 to fig 2b
%move 2c, 2e and 2f as "2d". These three figures should be without any label.
%Remove 2d.
The \fref{fig:fes_trpcage}a clearly indicates folded $(\mathbf F)$, semi-unfolded/intermediate $(\mathbf I)$ and unfolded $(\mathbf U)$ states of Trp-cage.
Representative structures of various conformational minima found on the free surface are obtained by clustering the trajectory of the corresponding GASS window (\fref{fig:fes_trpcage}f).
They indicate that the conformational space, including the unfolded states, is very well explored.
The free energy barrier for $\mathbf F \rightarrow \mathbf I$, and free energy difference between $\mathbf F$ and $\mathbf U$ are computed as a function of simulation time (\fref{fig:fes_trpcage}d).
This plot indicates that the estimate of free energy difference was converged by 100~ns and free energy barrier was converged by 60~ns.
%
%We interestingly find that the unfolded state is 0.2~kcal~mol$^{-1}$ lower than the folded state.
%
After 90~ns, new unfolded states were explored, resulting in a 
small drift in the convergence curve of $\Delta F$.
\changes{
%TODO-NN3
Detailed studies on the mechanism of folding have been done experimentally\cite{WountersenJPCB:2013,trp:2002:original,Qiu:2002,Ahmed:2005,Neuweiler:2005,Iavarone:2005,Bunagan:2006,Streicher:2007,Iavarone:2007,Mok:2007,Hudaky:2008,Barua:2008,Culik:2011,Rovo:2011,Ronge:2012,Anna:2012,Rovo:2013,ADAMS:2006,Tucker:JPCL:2020}
%20-35-references-of-WountersenJPCB:2013, refs-159-163-of-Ferguson:2018,
and computationally.\cite{WountersenJPCB:2013,Vijaypande:2002,Adrian:2002,CHOWDHURY:2003,Zhou:PNAS:2003,Bolhuis:PNAS:2006,Bolhuis:2008,HU:2008,Lindorff:2011,Shao:2012,WU:2011,Marino:2012,Lai:2013,XU:2008,Bandyopadhyay:PCCP:2016,Ferguson:2018,Kim:2016,Hatch:2014,Kannan:2014,Alexander:2005,Zhou:Proteins:2003,Levy:JPCB:2013,Patriksson:2007,Pitera:2003,trp:2002,be:mtd:1,be:mtd:2,trp:2011,trp:2012,trp:2013,trp:2014,trp:2015,trp:2015:amd,trp:2017,trp:2018,REHT:2021}
%36-48-references-of-WountersenJPCB:2013,refs-148-158-of-Ferguson:2018
%%TODO-NN3
The protein folding landscape of Trp-cage shows intermediate states, different to the classical  picture with only two states.
%%TODO-NN3
The presence of metastable intermediate states have been seen observed computationally~\cite{Zhou:PNAS:2003,Lindorff:2011} and experimentally.\cite{Sauer:PNAS:2005}
%%TODO-NN3
Two major folding pathways have been identified for this protein.\cite{Bolhuis:PNAS:2006, Debenedetti:JCP:2015,Laio:PLOS:2009,Levy:JPCB:2013}
The global minimum on the GASS free energy surface
(\fref{fig:fes_trpcage}a) is located at RMSD$=1.1$~{\AA},  
and Rg$=6.8$~{\AA}, and it corresponds to the folded state ($\mathbf{F}$).
The same was also observed in independent 120~ns WSMTD (\fref{fig:fes_trpcage}b) and 3~$\mu$s REST2 (\fref{fig:fes_trpcage}c) simulations.
The intermediate ($\mathbf{I}$) and the unfolded ($\mathbf{U}$) states are located on the (RMSD, Rg) free energy surface at ($5.7$~{\AA}, $7.5$~{\AA}), and ($6.2$~{\AA}, $11.0$~{\AA}), respectively.
The intermediate state {\bf I} and the unfolded state {\bf U} comprised of large number of protein conformations (\fref{fig:fes_trpcage}e).
The representative structures presented in \fref{fig:fes_trpcage}e were obtained by backbone RMSD based clustering of the biased GASS trajectories.
%
%for the windows corresponding to the 
%the broad basin of {\bf I}.
%
%Although we have not determined dominant the clusters and analyzed reaction pathways from the unbiased GASS trajectories, 
We also observed the intermediates {\bf SB-I}, {\bf LOOP-I}, and {\bf HLX-I} reported by Kim~et~al.\cite{Debenedetti:JCP:2015}; see Supporting Information.
However, it is noted that proper characterization of the intermediate states and reaction pathways require analysis of the unbiased GASS trajectories, which was not carried out here.
The free energy barrier $\mathbf{F}\rightarrow\mathbf{I}$ is $2.3$ kcal~mol$^{-1}$ using GASS, while it is  $2.4$ kcal~mol$^{-1}$ and $3.2$ kcal~mol$^{-1}$ using WSMTD and REST2 simulations, respectively.
From the GASS-computed free energy surface, we calculated the barrier for going from {\bf F} to {\bf U} state as $3.0$~kcal~mol$^{-1}$. 
The {\bf I} and the {\bf U} states are only $0.0$ and $0.7$~kcal~mol$^{-1}$ higher than the {\bf F} state, respectively.
%
%
 %$\mathbf{F}\rightarrow\mathbf{U}$ are $\approx 2.3$ kcal~mol$^{-1}$ 
%And the free energy barrier $\mathbf{F}\rightarrow\mathbf{U}$ is observed as ~$\sim 3.0$ kcal~mol$^{-1}$.
%TODO-NN3
The free energy data and the conformational landscape reported in our study are in excellent  qualitative and quantitative agreement with various reports in the literature.\cite{trp:2015:amd,Gracia:2010}
%Zhu:JCP:2012
%
Barrier for ${\bf F} \rightarrow {\bf I}$  
 also agrees with that reported in  Refs.\cite{REHT:2021,Lindorff:2011}
%TODO-NN3
The free energy difference between the unfolded and the folded state was experimentally measured as 0.77 kcal~mol$^{-1}$ at 298K by Streicher and Makhatadze,\cite{Makhatadze:BioChem:2007} which is in remarkably good agreement with our estimate ($0.7$~kcal~mol$^{-1}$).
%
%
%NEW REFS
%Ferguson:JCC:2018         https://doi.org/10.1002/jcc.25520
%Debenedetti:JCP:2015 https://doi.org/10.1063/1.4913322
%Bolhuis:PNAS:2006     https://doi.org/10.1073/pnas.0606692103 
%Makhatadze:BioChem:2007         https://doi.org/10.1021/bi602424x 
%WountersenJPCB:2013                https://doi.org/10.1021/jp404714c 
%Zhou:PNAS:2003                          https://doi.org/10.1073/pnas.2233312100 
%Bandyopadhyay:PCCP:2016        https://doi.org/10.1039/C6CP04634G 
%Gracia:2010                                  https://doi.org/10.1002/prot.22702
%Laio:PLOS:2009                            https://doi.org/10.1371/journal.pcbi.1000452 
%Levy:JPCB:2013                            https://doi.org/10.1021/jp401962k
%Tucker:JPCL:2020                         https://doi.org/10.1021/acs.jpclett.9b03706 
%Ferguson:2018                              https://doi.org/10.1002/jcc.25520
%Sauer:PNAS:2005.  https://doi.org/10.1073/pnas.0507351102 

%
Although the folded (${\mathbf{F}}$) and intermediate (${\mathbf{I}}$) states were reasonably sampled in WSMTD, the unfolded (${\mathbf{U}}$) states were not (\fref{fig:fes_trpcage}b).
The computed free energy difference between the {$\mathbf{U}$} and ${\mathbf{F}}$ states is  not converged in 120~ns of WSMTD (\fref{fig:fes_trpcage}d).
Sampling of the unfolded states was insufficient even after 3~$\mu$s of REST2; See \fref{fig:fes_trpcage}c.
The extent of exploration of the RMSD-Rg space is much less in REST2 compared to that in GASS.
Further, the free energy surface computed from REST2 trajectory is noisy.
Thus we think that REST2 free energy estimates are likely to be not converged.
These results underline the importance of the GASS approach.
}

%\newpage
\section{Conclusions}
It has been shown that the combination of the global tempering REST2 method with the CV-based biased sampling technique WSMTD, as done in the proposed GASS method, is an efficient way to study conformational sampling and compute free energies of large soft matter systems in solution.
A directed conformational sampling achieved by a restrained-bias along a CV, in tandem with an exhaustive sampling of orthogonal coordinates by MTD bias and REST2, makes the GASS method different to other sampling methods.
This is a much needed feature for studying various biochemical processes, such as protein folding.
Test calculations performed on solvated alanine tripeptide show that the GASS method provides accurate prediction of conformational free energy landscapes.
The method has been applied to study the unfolding/folding free energy landscape of Trp-cage protein in water, wherein a controlled unfolding of the protein was accomplished by applying restraining bias along the RMSD coordinate.
Orthogonal coordinates were enhanced sampled by REST2, in concert with the explicit MTD bias on Rg and $\alpha$-helicity CVs.
A quick convergence in free energy estimates was observed and a good conformational sampling of the unfolded states was noted.
The free energy landscape projected on the RMSD-Rg space has three distinct free energy minima corresponding to fully folded, partially unfolded (intermediate), and unfolded states.
On the landscape, the free energy barrier to go from folded to the intermediate state is $2.3$~kcal~mol$^{-1}$ and the folded to the unfolded state is $3.0$~kcal~mol$^{-1}$. 
The intermediate and the unfolded states are only $0.0$ and $0.7$~kcal~mol$^{-1}$ higher than the folded state, respectively.
We hope that the new method proposed here will be very useful to study mechanism and free energetics of complex biochemical processes such as protein folding, drug-binding, and  diffusion of molecules through membrane.
%{\textcolor{red}{think}}
%
\changes{The reweighting codes and the input files for performing GASS simulations reported here are  available online.\cite{github:gass} }

\begin{acknowledgments}
Authors gratefully acknowledge the discussions with Prof. Ricardo L. Mancera (Curtin University). ABK thanks IIT Kanpur and Curtin University for the PhD fellowship and travel support. Authors thank IIT Kanpur for providing computing resources at the HPC2013 cluster.
%
% in understanding the Tiwary--Parrinello reweighting scheme. 
%Authors also thank IIT Kanpur for availing the computing resources. 
%NN acknowledges IIT Kanpur for the support through P.~K. Kelkar Young Faculty Research Fellowship. 
%SA thanks UGC for Ph.~D fellowship.
\end{acknowledgments}

\section*{Supporting Information} 
Additional supporting information may be found in the online version of this article. 

%\bibliographystyle{jcc}
%\bibliography{references.bib}

\end{document}